\documentclass{lhep}

\vol{X} \jyear{XXXX} \pages{XXX-XXX}

\received{28 September 2018}
\published{26 November 2018}

\def\be{\begin{equation}}
\def\ee{\end{equation}}
\def\bea{\begin{eqnarray}}
\def\eea{\end{eqnarray}}

\begin{document}

\title{Fermionic Dark Matter and New Scalar Production in
	$e^+e^- \to H^+H^-$ at Colliders}

\author{\textbf{Asmaa AlMellah}$^{1,*}$, \textbf{Faeq Abed}$^{2,*}$, \textbf{Gaber Faisel}$^{1}$}
\address{
	$^{1}$ Department of Physics, Faculty of Arts and Sciences, Suleyman Demirel University, 32260, Isparta, Turkey.\\
	$^{}$Corresponding Author: Email: asmaa.almellah@gmail.com - ORCID: 0000-0003-3975-3504,\\
	$^{}$Email: gaber.faisel@sdu.edu.tr - ORCID: 0000-0001-8770-1966.\\
	$^{2}$Iraqi Nuclear Regulatory Authority, Baghdad, Iraq.\\
	$^{}$Corresponding Author: Email: faeq.abed@irsra.gov.iq - ORCID: 0000-0002-3934-4367}	
\begin{abstract}
	We investigate the pair production process $e^+e^- \to H^+H^-$ in the framework of the scotogenic model. The production mechanism receives contributions at tree level from photon and $Z$-boson exchange, as well as from $t$-channel exchange of the new singlet right-handed fermions $N_{1,2,3}$. where neutrino masses are generated radiatively and one of the singlet right-handed fermions serves as a viable dark matter candidate. We evaluate the individual contributions of these diagrams and compute the total production cross section after imposing all relevant theoretical and experimental constraints on the model parameters, including those associated with dark matter relic abundance and direct detection limits. Our results demonstrate that the dominant contribution to the cross section originates from the exchange of the singlet fermions $N_{1,2,3}$, particularly from the dark matter component of the spectrum. In addition, we examine the dependence of the cross section on the center-of-mass energy for several benchmark scenarios in the allowed parameter space. These predictions can be probed at future high-energy $e^+e^-$ colliders, providing a sensitive test of the scotogenic framework and the role of fermionic dark matter, as well as enabling more stringent constraints on the model parameters.
\end{abstract}

\maketitle


\section{{ Introduction }}

The requirement for the existence of New Physics (NP) beyond the Standard Model (SM) arises from several fundamental limitations of the SM. Notably, the SM lacks a mechanism for generating neutrino masses and fails to provide a viable candidate for Dark Matter (DM). Additional shortcomings include the absence of gravity within the SM framework and its inability to account for the baryon asymmetry observed in the Universe. These deficiencies collectively highlight the need for extensions to the SM.

One of the prominent extensions of the Standard Model (SM) is the scotogenic model, introduced by Ma in 2006 [1]. This model offers a compelling mechanism for generating small neutrino masses, which is consistent with experimental observations. In addition to addressing the neutrino mass problem, the scotogenic model provides a viable candidate for dark matter (DM). Specifically, the DM candidate can originate from the new particles introduced by the model, including components of the scalar doublet \( \eta \) or the lightest of the three singlet Majorana fermions \( N_{1,2,3} \). These additional particles, \( \eta \) and \( N_i \), are proposed extensions to the particle content of the SM. A crucial feature of the model is the imposition of a \( Z_2 \) symmetry, under which all SM particles are \( Z_2 \)-even, while the newly introduced particles are \( Z_2 \)-odd. This symmetry ensures the stability of the DM candidate, further enhancing the model's consistency and appeal. 

In Ref. \cite{Ho:2013spa}, the process \( e^+e^- \to H^+H^- \to \ell^+\ell^{\prime-}\, \slash\!\!\!\!\!E \) was investigated for potential collider signatures. With advancements over the past decade in neutrino oscillation experiments and updated constraints on neutrino masses derived from cosmological observations, it is now possible to revise and refine the constraints presented in that study. Additionally, recent measurements of the dark matter relic density provide an essential new dataset to incorporate into the analysis, offering the potential for stronger constraints on the model.  

Updating these constraints is crucial for accurately predicting the cross section of the process \( e^+e^- \to H^+H^- \), which was not analyzed in Ref. \cite{Ho:2013spa}. This process is significant as the produced \( H^+H^- \) pair can decay into various final states with different particle combinations, all of which depend on the magnitude of the \( e^+e^- \to H^+H^- \) cross section. Therefore, a thorough analysis of this cross section is essential.  

In this study, we perform such an analysis, focusing on the individual contributions to the cross section from \( Z \)-boson exchange, photon exchange, and the new singlet fermions introduced by the scotogenic model. This comprehensive approach provides a more complete understanding of the process and its implications within the model.

\section{{ The scotogenic model}}

The scotogenic model extends the Standard Model (SM) scalar sector by introducing an additional scalar doublet, \(\eta\). The scalar sector of the scotogenic model is described by the following Lagrangian:  
\begin{eqnarray} \label{L}
	{\cal L} = ({\cal D}^\mu \Phi)^\dagger\,{\cal D}_\mu \Phi \,+\,
	({\cal D}^\mu\eta)^\dagger\,{\cal D}_\mu\eta \;-\; {\cal V} ~,
\end{eqnarray}
where \(\Phi\) represents the SM Higgs doublet, \(\mathcal{D}_\mu\) is the covariant derivative that includes SM gauge fields, and \(\mathcal{V}\) denotes the scalar potential, given by~\cite{Ma:2006km}:  
\begin{eqnarray*} \label{potential}
	&&{\cal V} =\mu_1^2\,\Phi^\dagger \Phi+ \mu_2^2\,\eta^\dagger\eta
	+\mbox{$\frac{1}{2}$}\lambda_1^{}(\Phi^\dagger \Phi)^2 +
	\mbox{$\frac{1}{2}$}\lambda_2^{}(\eta^\dagger\eta)^2 \nonumber \\
	&& +\; \lambda_3^{}(\Phi^\dagger \Phi)(\eta^\dagger\eta) \,+\,
	\lambda_4^{}(\Phi^\dagger\eta)(\eta^\dagger \Phi) \,+\,
	\mbox{$\frac{1}{2}$}\lambda_5^{}\bigl[
	(\Phi^\dagger\eta)^2+(\eta^\dagger\Phi)^2\bigr] ~,
\end{eqnarray*}
After electroweak symmetry breaking, the fields are expressed as:  
\begin{eqnarray}
	\Phi \,\,=\, \left(\!\begin{array}{c} 0 \vspace{1pt} \\
		\frac{1}{\sqrt2}_{\vphantom{o}}(h+v) \end{array}\! \right) ,
	\hspace{7ex}
	\eta \,\,=\, \left(\!\begin{array}{c} H^+ \vspace{1pt} \\
		\frac{1}{\sqrt2}_{\vphantom{o}}({\cal S}+i{\cal P}) \end{array}\!
	\right) ,\label{phi}
\end{eqnarray}
where \(h\) is the physical Higgs boson, and \(v\) is the vacuum expectation value (VEV) of \(\Phi\). Due to the imposed \(Z_2\) symmetry, the VEV of \(\eta\) is zero.  

The scalar masses are given by:  
\[
m_\mathcal{S}^2 = m_\mathcal{P}^2 + \lambda_5 v^2 = \mu_2^2 + \frac{1}{2} (\lambda_3 + \lambda_4 + \lambda_5) v^2,
\]
\[
m_{H^\pm}^2 = \mu_2^2 + \frac{1}{2} \lambda_3 v^2.
\]
In the limit of \(|\lambda_5| \ll |\lambda_3 + \lambda_4|\), it follows that \(m_\mathcal{S}^2 \simeq m_\mathcal{P}^2\)~\cite{Kubo:2006yx}.  

The interactions and masses of the new singlet Majorana fermions (\(N_k\)) are described by the Lagrangian:  
\begin{eqnarray} \label{LN}
	{\cal L}_N^{} =-\mbox{$\frac{1}{2}$} M_k^{}\,\overline{N_k^{\rm
			c}}\,P_R^{} N_k^{} \,+\, {\cal Y}_{rk}^{} \Bigl[ \bar\ell_r^{} H^-
	- \mbox{$\frac{1}{\sqrt2}$}\bar\nu_r^{}({\cal S}-i {\cal P})
	\Bigr] P_R^{} N_k^{} +{\rm H.c.} ~,
\end{eqnarray}
where \(\ell_{1,2,3} = e, \mu, \tau\), \(\mathcal{Y}_{rk}\) are the Yukawa couplings, \(M_k\) represents the masses of \(N_k\), \(P_R = \frac{1}{2}(1+\gamma_5)\), and \(k, r = 1, 2, 3\).  

The interactions of \(H^\pm\) with the photon (\(A\)) and \(Z\)-boson are derived from the Lagrangian:
\begin{eqnarray} \label{Lphieta}
	{\cal L}_H^\pm &\,\supset&\, \; i e\,\bigl(H^+\,\partial^\rho
	H^--H^-\,\partial^\rho H^+\bigr) A_\rho^{} \,\nonumber
	\\ && +\; \frac{g}{2c_{\rm w}} \bigl[ \, i\bigl(1-2s_{\rm
		w}^2\bigr) \bigl(H^+\,\partial^\rho H^--H^-\,\partial^\rho
	H^+\bigr) \bigr] Z_\rho^{} ~,
\end{eqnarray}
where \(e = g s_{\text{w}}\), \(c_{\text{w}} = \cos\theta_{\text{W}}\), and \(s_{\text{w}} = \sin\theta_{\text{W}}\), with \(\theta_{\text{W}}\) being the Weinberg angle.  

This framework establishes the key scalar and fermionic properties and interactions in the scotogenic model.

\subsection{Neutrino masses generation }

The imposed $Z_2$ symmetry in the scotogenic model prevents neutrino masses at the tree level. However, at the one-loop level, neutrino masses are generated through the mediation of $\cal S$, $\cal P$, and $N_k$. The eigenvalues of the light neutrino masses $m_i^{}$ are determined by the loop-induced quantity~\cite{Ma:2006km}:

\begin{eqnarray}
	\Lambda_k^{} \,\,=\,\, \frac{\lambda_{5\,}^{}v^2}{16\pi^2M_k^{}} 
	\Biggl[ \frac{M_k^2}{m_0^2-M_k^2} + \frac{2 M_k^4\,\ln\bigl(M_k^{}/m_0^{}\bigr)}
	{\bigl(m_0^2-M_k^2\bigr)\raisebox{1.7pt}{$^{\!2}$}} \Biggr]^{\vphantom{\int^|}} ~,
\end{eqnarray}

where $m_0=\frac{1}{2}\big(m_{\cal S}+m_{\cal P}\big)\simeq m_{\cal S}^{}\simeq m_{\cal P}^{}$. The neutrino mass matrix is expressed as:

\begin{eqnarray}
	\displaystyle {\cal M}_\nu^{} \,\,= \,\, Y\, {\rm diag}(\Lambda_1,\Lambda_2,\Lambda_3)\, Y^{{\rm T}^{\vphantom{\displaystyle|}}} \,, \label{lambdak}
\end{eqnarray}

and it can be diagonalized by:

\begin{eqnarray} 
	\displaystyle {\rm diag}\bigl(m_1^{},m_2^{},m_3^{}\bigr) \,\,=\,\, {\cal U}^\dagger{\cal M}_\nu\,{\cal U}^* \,,  \label{Mnu}
\end{eqnarray}

where the unitary matrix $\cal U$ represents the Pontecorvo-Maki-Nakagawa-Sakata (PMNS) matrix. In our analysis, the PDG parametrization~\cite{pdg} is utilized, given by $ \displaystyle {\cal U} = \tilde u\;{\rm diag}\bigl(e^{i\alpha_1/2},e^{i\alpha_2/2},1\bigr)$, with $\alpha_{1,2}^{}$ being the Majorana $CP$-violation phases, and $\tilde u$ defined in terms of $c_{mn}^{}=\cos\theta_{mn}^{}\ge0$, $s_{mn}^{}=\sin\theta_{mn}^{}\ge0$, and the Dirac phase $\delta$. The matrix $\tilde u$ is detailed in Eq.~(10) of Ref.~\cite{Faisel:2014gda}. Analytical solutions for Eqs.~(\ref{UMU}) are given by~\cite{Ho:2013spa}:

\begin{eqnarray} \label{123} 
	Y_{e1}^{} &=& \frac{-c_{12\,}^{}c_{13}^{}\,Y_1^{}}
	{c_{12\,}^{}c_{23\,}^{}s_{13\,}^{}e^{i\delta}-s_{12\,}^{}s_{23}^{}}, \,
	Y_{e2}^{} = \frac{-s_{12\,}^{}c_{13}^{}\,Y_2^{\vphantom{\int}}}
	{s_{12\,}^{}c_{23\,}^{}s_{13\,}^{}e^{i\delta}+c_{12\,}^{}s_{23}^{}}, \nonumber\\
	Y_{\mu1}^{} &=& \frac{\big(c_{12\,}^{}s_{23\,}^{}s_{13\,}^{}e^{i\delta}+s_{12\,}^{}c_{23}^{}\big)Y_1^{}}
	{c_{12\,}^{}c_{23\,}^{}s_{13\,}^{}e^{i\delta}-s_{12\,}^{}s_{23}^{}}\, ,\,
	Y_{\mu3}^{} = \frac{s_{23\,}^{}Y_3^{\vphantom{\int}}}{c_{23}^{}} , \nonumber \\
	Y_{e3}^{} &=& \frac{s_{13}^{}\,Y_3^{}}
	{c_{23\,}^{}c_{13\,}^{}e^{i\delta}}\, , \,
	Y_{\mu2}^{} = \frac{\big(s_{12\,}^{}s_{23\,}^{}s_{13\,}^{}e^{i\delta}-c_{12\,}^{}c_{23}^{}\big)Y_2^{}}
	{s_{12\,}^{}c_{23\,}^{}s_{13\,}^{}e^{i\delta}+c_{12\,}^{}s_{23}^{}}, \nonumber \\
	Y_{\tau k}^{} &=& Y_{k}^{} \; ~,
\end{eqnarray}

which correspond to the light neutrino mass eigenvalues~\cite{Ho:2013spa}:

\begin{eqnarray} \label{m1m2m3}
	m_1^{} \,=\,
	\frac{\Lambda_{1\,}^{}Y_{e1}^2\,e^{-i\alpha_1}}{c_{12\,}^2c_{13}^2}~, \,
	m_2^{} \,=\,
	\frac{\Lambda_{2\,}^{}Y_{e2}^2\,e^{-i\alpha_2}}{s_{12\,}^2c_{13}^2}~, \,
	m_3^{} \,=\,
	\frac{\Lambda_{3\,}^{}Y_3^2}{c_{13\,}^2 c_{23}^2} ~.
\end{eqnarray}

The Majorana $CP$-violation phases are determined using the relations:

\begin{eqnarray} \label{alpha12} 
	\displaystyle \alpha_1^{} = \arg\bigl(\Lambda_{1\,}^{}Y_{e1}^2\bigr)~, \,
	\alpha_2^{} = \arg\bigl(\Lambda_{2\,}^{}Y_{e2}^2\bigr)~, \,
	\arg\bigl(\Lambda_{3\,}^{}Y_3^2\bigr) \,\,=\,\, 0~.
\end{eqnarray}

\subsection{Dark Matter}

The scotogenic model extends the Standard Model (SM) by introducing additional particles in the fermionic and scalar sectors, specifically $N_{1,2,3}$ and $S, P, H^\pm$, respectively. These new particles are odd under the $Z_2$ symmetry, ensuring that the lightest among them is stable and can serve as a candidate for dark matter (DM).

In this study, we focus on the commonly considered scenario where $N_1$ acts as the DM particle, with $N_2$ nearly degenerate in mass with $N_1$. This configuration is advantageous because it simultaneously satisfies the constraints from the DM relic density and the branching ratio (BR) of the lepton flavor-violating process $\mu \to e \gamma$.

The relic density is represented in terms of the current DM density relative to its critical value, $\Omega$, and the Hubble parameter, $\hat h$, as $\Omega \hat h^2$. Theoretical estimation of this quantity follows the expression ~\cite{Kubo:2006yx,Jungman:1995df}:

\begin{eqnarray} \label{omega}
	\Omega \hat h^2 &\,=\,& \frac{1.07\times10^9\;x_f^{}{\rm\;GeV}^{-1}}{\sqrt{g_*^{}}\;m_{\rm Pl}^{}\,\bigl[a_{eff}+3(b_{eff}-a_{eff}/4)/x_f^{}\bigr]} ~,\nonumber \\ 
	x_f^{} &\,=\,& \ln\frac{0.191\,\bigl(a_{eff}+6 b_{eff}/x_f^{}\bigr)M_1^{}\,m_{\rm Pl}^{}}{\sqrt{g_*^{}\,x_f^{}}} ~
\end{eqnarray}

Here, $g_*^{}$ is the effective number of relativistic degrees of freedom below the freeze-out temperature $T_f^{}=M_1^{}/x_f^{}$, and $m_{\rm Pl}^{}=1.22\times10^{19}$\,GeV is the Planck mass. The parameters $a_{eff}$ and $b_{eff}$ are derived from the expansion of the coannihilation rate, $\sigma_{eff} v_{\rm rel}^{} = a_{eff} + b_{eff} v_{\rm rel}^2$, in terms of the relative speed $v_{\rm rel}$ of the annihilating particles in their center-of-mass frame. The effective cross-section is given as \,$\sigma_{eff} = \frac{1}{4}\big( \sigma_{11} + \sigma_{12} + \sigma_{21} + \sigma_{22} \big)$, where the individual cross sections $\sigma_{ij}$ (for $i,j=1,2$) are expressed as:

\be \sigma_{ij} = \sigma_{N_i^{} N_j^{} \to \ell_i^- \ell_j^+} + \sigma_{N_i^{} N_j^{} \to \nu_i^{} \nu_j^{}} \ee

The above cross sections have been computed in Refs.\cite{Ho:2013hia,Ho:2013spa} and arise from $t$- and $u$-channel tree-level diagrams mediated by $H^\pm$ and $({\cal S, P})$, depending on whether the final states are charged leptons or neutral neutrinos.

In our analysis, to constrain the parameter space of the model using the observed DM relic density, we assume that the DM and the new scalar particles are not degenerate in mass. This avoids the contributions of scalar co-annihilation processes to the relic density.

\subsection{ $e^+(p_+)_{\,}e^-(p_-)\to H^+H^-$}

In the scotogenic model considered in this study, the amplitude for the process ,$e^+(p_+){,}e^-(p-)\to H^+H^-$, receives contributions from tree-level diagrams involving the exchange of a photon ($\gamma$), a $Z$ boson, and $N_{1,2,3}$. Assuming massless $e^\pm$, the resulting cross section is expressed as \cite{Ho:2013spa}.
\begin{eqnarray} \label{sigma_ee2HH}
\sigma_{e^+e^-\to H^+ H^-} = \frac{\pi\alpha^2\beta^3}{3 s} +\frac{\alpha}{12} \frac{\bigl(g_L^2+g_L^{}g_R^{}\bigr)\beta^3}{s-m_Z^2} +
    \frac{\bigl(g_L^4+g_L^2g_R^2\bigr)\beta^3s}{96\pi\bigl(s-m_Z^2\bigr)\raisebox{1.1pt}{$^{\!2}$}}
     \nonumber \\ +
    \raisebox{0.5ex}{\footnotesize$\displaystyle\sum_k$}\, \frac{|{\cal Y}_{1k}|^4}{64\pi\,s}
    \Biggl(w_k^{}\,\ln\frac{w_k^{}+\beta}{w_k^{}-\beta}-2\beta\Biggr)
    \! +\;
    \Biggl[ \frac{\alpha}{16 s} + \frac{g_L^2}{64\pi\,\bigl(s-m_Z^2\bigr)}\Biggr]\nonumber\\
    \raisebox{0.5ex}{\footnotesize$\displaystyle\sum_k$}\,|{\cal Y}_{1k}|^2
    \biggl[\bigl(w_k^2-\beta^2\bigr)\,\ln\frac{w_k^{}+\beta}{w_k^{}-\beta}-2\beta\,w_k^{}\biggr]
    \nonumber \\  \! +\;
    \raisebox{0.5ex}{\footnotesize$\displaystyle\sum_{j,\,k>j}$}\,
    \frac{|{\cal Y}_{1j\,}{\cal Y}_{1k}|^2}{64\pi\,s}
    \Biggl( \frac{w_j^2-\beta^2}{w_j^{}-w_k^{}}~\ln\frac{w_j^{}+\beta}{w_j^{}-\beta}
    + \frac{w_k^2-\beta^2}{w_k^{}-w_j^{}}~\ln\frac{w_k^{}+\beta}{w_k^{}-\beta}\nonumber \\  - 2\beta \Biggr)
\end{eqnarray}
here \,$j,k=1,2,3$, \,$s=(p_++p_-)^2$,\, $\alpha \,\,=\,\,
\frac{e^2}{4\pi}$, $\beta \,\,= \sqrt{1-\frac{4m_H^2}{s}}$ and $
w_k^{} \,\,=\,\, 1 \,+\, \frac{2M_k^2}{s} \,-\, \frac{2m_H^2}{s}
\beta$. In the numerical analysis, we employ the effective values
\,$\alpha=1/128$,\, $g=0.6517$,\, and \,$s_{\rm
w}^2=0.23146$\,~\cite{pdg}.

\subsection{Constraints}

In this study, we adopt the normal ordering (NO) of neutrino masses. The mixing angles, Dirac phase, $|\Delta m^2_{31}|$, and $\Delta m^2_{21}$ are determined from various measurements. For numerical evaluation, we use the results of the global neutrino oscillation data fit presented in Ref.~\cite{deSalas:2020pgw}. This fit imposes the constraint $32.0 \,\,<\,\, R_m\equiv \frac{|\Delta m^2_{31}|}{\Delta m^2_{21}} \,\,<\,\, 36.0$ on the parameter space, based on the 90\% CL ranges of the data. Additionally, measurements from the Planck satellite, BAO observations, H(z) data, and Supernovae Ia impose a stringent $2\sigma$ upper limit of $\sum m_i < 0.12$ eV. Constraints from neutrinoless double beta decay experiments yield $|\langle m\rangle_{ee}| < 0.06 - 0.2$ eV at the (95 hundred percent) confidence level~\cite{EXO-200:2019rkq,KamLAND-Zen:2016pfg,GERDA:2020xhi}, where $\langle m\rangle_{ee}$ is defined as $\langle m\rangle_{ee}= m_1 U^2_{e1}+ m_2 U^2_{e2}+m_3 U^2_{e3}$.

The Yukawa interactions, which include the charged Higgs $H^\pm$ as shown in Eq.(\ref{LN}), generate one-loop diagrams contributing to lepton flavor violation (LFV) processes. Detailed discussions of these processes are provided in Ref.~\cite{Toma:2013zsa}. Currently, experimental upper bounds on the branching ratios of such processes are BR$(\mu\to e\gamma)<4.2\times10^{-13}$~\cite{MEG:2016leq}, BR$(\tau\to e\gamma)<3.3\times10^{-8}$~\cite{BaBar:2009hkt}, and BR$(\tau\to \mu\gamma)<4.4\times10^{-8}$~\cite{BaBar:2009hkt}, with the most stringent constraint arising from BR$(\mu\to e\gamma)$. The expressions for these branching ratios in the scotogenic model are available in Ref.~\cite{Toma:2013zsa}. Meanwhile, the flavor-diagonal counterparts of these LFV processes modify the anomalous magnetic moment $a_{\ell_i}$ as~\cite{Ma:2001mr}:

\begin{eqnarray} \label{g-2}
	\Delta a_{\ell_i}^{} \,\,=\,\,
	\frac{-m_{\ell_i}^2}{16\pi^2m_H^2}\,\sum_k\,|Y_{ik}|^2\, {\cal
		F}\bigl(M_k^2/m_H^2\bigr) ~.
\end{eqnarray}

It is observed that the anomalous magnetic moment of the muon provides a stronger bound on the scotogenic parameter space compared to those from the electron and tau. The difference between the SM prediction and the current experimental value for $a_\mu^{}$ is $a_\mu^{\rm exp}-a_\mu^{\rm SM}=(2.51\pm 0.59)\times10^{-9}$~\cite{Aoyama:2020ynm}.

Direct detection of dark matter (DM) via the interaction of $N_1$ with nucleons, mediated by Higgs exchange at one-loop level, is discussed in Ref.~\cite{Ibarra:2016dlb}. To evade stringent constraints from direct detection~\cite{XENON:2018voc,PandaX-4T:2021bab}, we follow Ref.~\cite{Ibarra:2016dlb,Liu:2022byu} and set $\lambda_{3,4}= 0.01$. Consequently, $m_{0} \simeq m_{H^\pm} + \frac{1}{2} \lambda_{4} v^2 \simeq m_{H^\pm}+350$ GeV. Notably, direct detection bounds were not accounted for in Refs.~\cite{Ho:2013spa}. However, in this study, these bounds are included alongside the latest global neutrino oscillation data from Ref.~\cite{deSalas:2020pgw}.

In the following section, we will present our results and analyze the implications. First, the mixing angles $\theta_{12,23,13}$ and Dirac phase $\delta$ are fixed to their central values based on the global neutrino oscillation data in Ref.~\cite{deSalas:2020pgw}. Subsequently, a scan is performed over the model's parameter space, including the masses of the new scalars $m_H, m_0$, the singlet fermions $N_{1,2,3}$, and the input parameters $Y_{1,2,3}$ appearing in the Yukawa couplings listed in Eq.(\ref{123}). For light DM masses $M_1<100$ GeV, LFV processes~\cite{Vicente:2014wga} and LHC searches~\cite{ATLAS:2019lng,CMS:2020bfa} predominantly exclude the parameter space. Concerning scalar masses, data on $W$ and $Z$ widths and the absence of direct evidence for new particles at $e^+e^-$ colliders impose the following upper bounds~\cite{Arhrib:2012ia,Cao:2007rm,Pierce:2007ut}:

\begin{eqnarray}
	m_{H^\pm} +m_{{\cal S},{\cal P}} > m_{W^\pm}, m_{H^\pm} > 70
	{\rm\;GeV}, m_{\cal S}^{}+m_{\cal P}^{} >\ m_Z .
\end{eqnarray}

One should remark that there are constrains on the
charged Higgs mass from the $\beta\rightarrow \tau \nu$,$\beta\rightarrow s\gamma$ and
from the direct measurements of the charged Higgs
decays at the LHC. In our analysis, we take into
account all these constraints. 

\section{Results and discussions}

This study considers the normal ordering (NO) of neutrino masses. The mixing angles, Dirac phase, $|\Delta m^2_{31}|$, and $\Delta m^2_{21}$ are constrained by various measurements. For numerical evaluations, the global neutrino oscillation data fit from Ref.~\cite{deSalas:2020pgw} is used. The fit imposes the condition $32.0 \,\,<\,\, R_m\equiv \frac{|\Delta m^2_{31}|}{\Delta m^2_{21}} \,\,<\,\, 36.0$ at the 90\% CL. Furthermore, data from the Planck satellite, BAO observations, H(z) measurements, and Supernovae Ia restrict the sum of neutrino masses to $\sum m_i < 0.12$ eV at $2\sigma$. Neutrinoless double beta decay experiments constrain $|\langle m\rangle_{ee}|$ to the range $0.06 - 0.2$ eV (95\% CL)~\cite{EXO-200:2019rkq,KamLAND-Zen:2016pfg,GERDA:2020xhi}, where $\langle m\rangle_{ee}= m_1 U^2_{e1}+ m_2 U^2_{e2}+m_3 U^2_{e3}$.  

The Yukawa interactions involving the charged Higgs $H^\pm$, as described in Eq.(\ref{LN}), generate one-loop diagrams contributing to lepton flavor violation (LFV) processes. These are discussed in Ref.~\cite{Toma:2013zsa}. Current experimental upper bounds on the branching ratios of such processes include BR$(\mu\to e\gamma)<4.2\times10^{-13}$~\cite{MEG:2016leq}, BR$(\tau\to e\gamma)<3.3\times10^{-8}$~\cite{BaBar:2009hkt}, and BR$(\tau\to \mu\gamma)<4.4\times10^{-8}$~\cite{BaBar:2009hkt}, with the most stringent constraint arising from BR$(\mu\to e\gamma)$. Branching ratio expressions in the scotogenic model are detailed in Ref.~\cite{Toma:2013zsa}. The flavor-diagonal counterparts of these LFV processes alter the anomalous magnetic moment $a_{\ell_i}$ as~\cite{Ma:2001mr}:  

\begin{eqnarray} \label{g-2}
	\Delta a_{\ell_i}^{} \,\,=\,\,
	\frac{-m_{\ell_i}^2}{16\pi^2m_H^2}\,\sum_k\,|Y_{ik}|^2\, {\cal
		F}\bigl(M_k^2/m_H^2\bigr) ~.
\end{eqnarray}

The muon anomalous magnetic moment places stronger constraints on the scotogenic parameter space than the electron or tau. The difference between the experimental and SM values is $a_\mu^{\rm exp}-a_\mu^{\rm SM}=(2.51\pm 0.59)\times10^{-9}$~\cite{Aoyama:2020ynm}.  

Direct dark matter (DM) detection, mediated by Higgs exchange at the one-loop level, is discussed in Ref.~\cite{Ibarra:2016dlb}. To avoid stringent detection constraints~\cite{XENON:2018voc,PandaX-4T:2021bab}, we adopt $\lambda_{3,4}= 0.01$, following Refs.~\cite{Ibarra:2016dlb,Liu:2022byu}, leading to $m_{0} \simeq m_{H^\pm} + \frac{1}{2} \lambda_{4} v^2 \simeq m_{H^\pm}+350$ GeV. While previous works, such as Ref.~\cite{Ho:2013spa}, did not consider direct detection bounds, they are included here along with the latest global neutrino oscillation data~\cite{deSalas:2020pgw}.  

The results and implications are presented next. The mixing angles $\theta_{12,23,13}$ and Dirac phase $\delta$ are fixed to their central values from Ref.~\cite{deSalas:2020pgw}. A scan over the parameter space includes scalar masses $m_H, m_0$, singlet fermions $N_{1,2,3}$, and Yukawa parameters $Y_{1,2,3}$ from Eq.(\ref{123}). For $M_1<100$ GeV, LFV processes~\cite{Vicente:2014wga} and LHC searches~\cite{ATLAS:2019lng,CMS:2020bfa} largely exclude the parameter space. Constraints on scalar masses from $W$ and $Z$ decays and $e^+e^-$ collider data impose limits~\cite{Arhrib:2012ia,Cao:2007rm,Pierce:2007ut}:  

\begin{eqnarray}
	m_{H^\pm} +m_{{\cal S},{\cal P}} > m_{W^\pm}, m_{H^\pm} > 70
	{\rm\;GeV}, m_{\cal S}^{}+m_{\cal P}^{} >\ m_Z .
\end{eqnarray}

\begin{figure}[h!]
	\centering
	\includegraphics[height=1.5in]{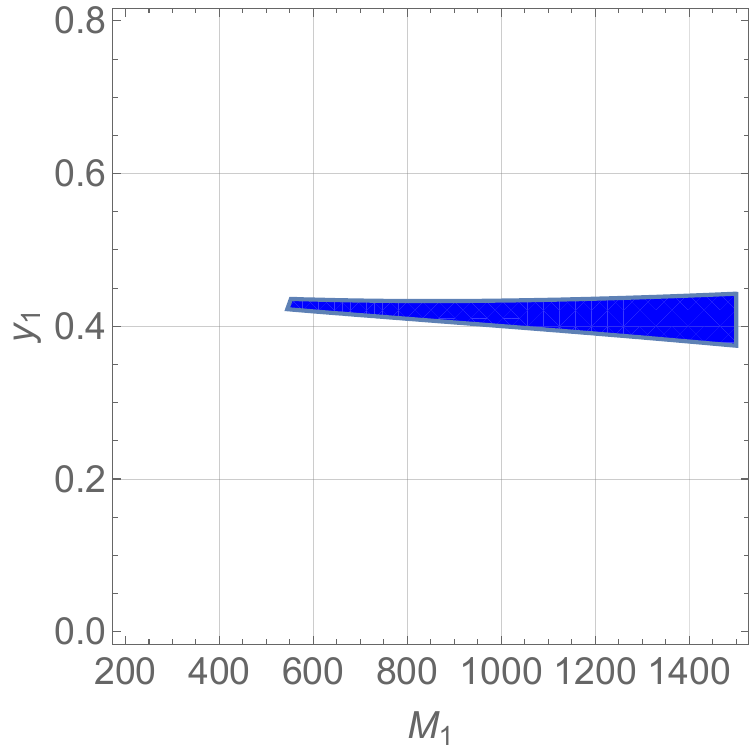}
	\caption{Region in magenta color is allowed by $\mu\to e\gamma$
		constraint  for the parameters $M_2=M_1$, $M_3=M_1+380$ GeV, $m_H
		= M_1 +400$ GeV, $Y_2= 0.49$ and $Y_3 = 0.66$.} \label{f:dental}
\end{figure}

\begin{figure}[h!]
	\centering
	\includegraphics[height=1.5in]{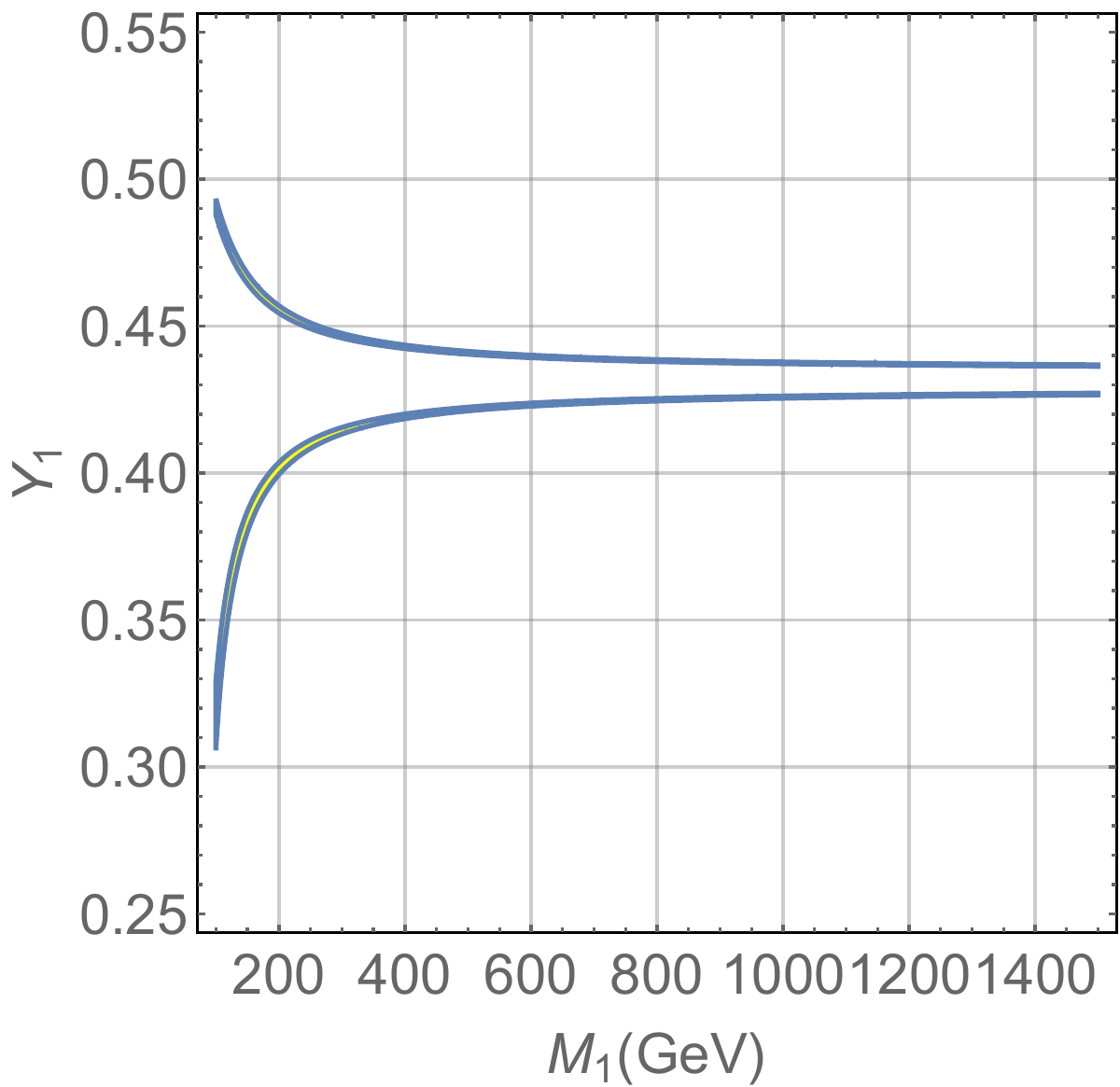}
	\caption{Allowed regions after imposing $\frac{|\Delta
			m^2_{31}|}{\Delta m^2_{21}}$ constraints  for the choice of the
		parameters as $M_2=M_1$, $M_3=M_1+380$ GeV, $m_0 = M_1 +750$ GeV,
		$Y_2= 0.49$ and $Y_3 = 0.66$.} \label{f:denta2}
\end{figure}

\begin{figure}[h!]
	\centering
	\includegraphics[height=1.55in]{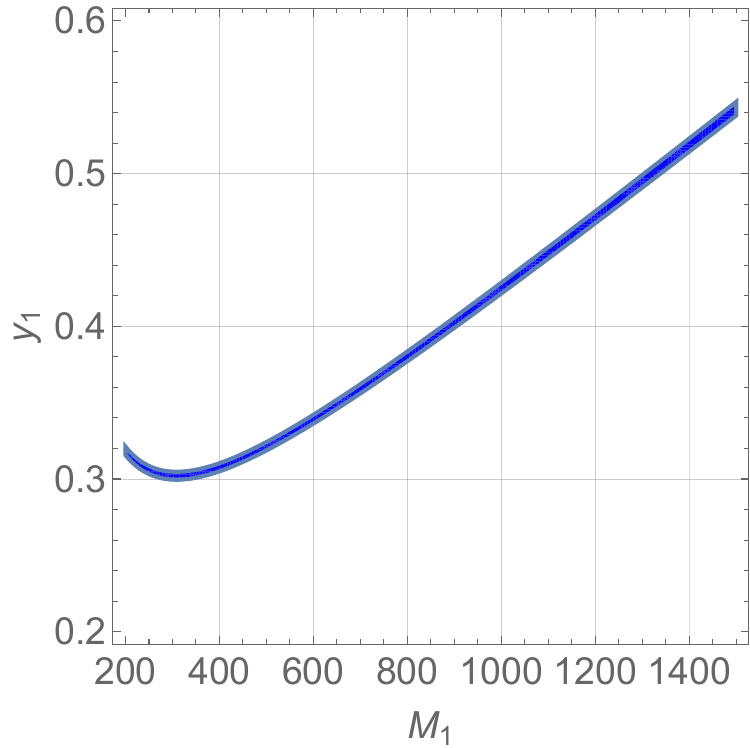}
	\caption{Allowed regions in the $M_1-Y_1$ plane  by $\Omega \hat
		h^2$ constraints for the parameters $M_2=M_1$, $M_3=M_1+380$ GeV,
		$m_H = M_1 +400$ GeV, $m_0 = M_1 +750$ GeV, $Y_2= 0.49$ and $Y_3
		= 0.66$. }\label{f:dental3}
\end{figure}

\begin{figure}[h!]
	\centering
	\includegraphics[height=1.55in]{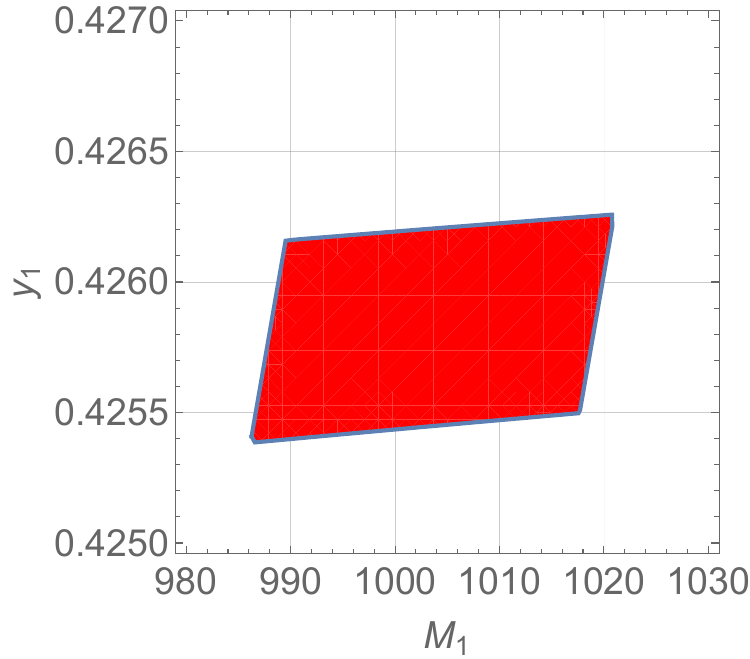}
	\caption{Allowed regions in the $M_1-Y_1$ plane  by $\mu\to
		e\gamma$, $\frac{|\Delta m^2_{31}|}{\Delta m^2_{21}}$, $\Omega
		\hat h^2$ constraints $M_2=M_1$, $M_3=M_1+380$ GeV,  $m_H = M_1
		+400$ GeV, $m_0 = M_1 +750$ GeV, $Y_2= 0.49$ and $Y_3 = 0.66$.
	}\label{f:dental4}
\end{figure}

\begin{figure}[h!]
	\centering
	\includegraphics[height=1.55in]{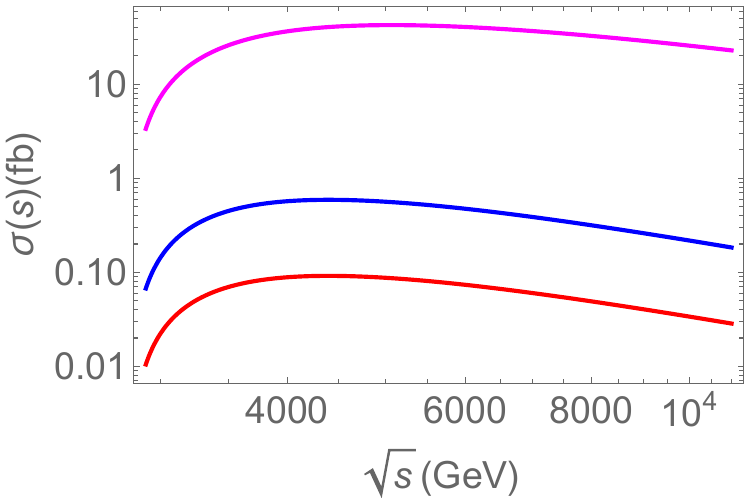}
	\caption{ Z, photon and new singlet fermions $N_{1,2,3}$
		individual contributions to the cross section of $\sigma_{e^+
			e^-\to H^+ H^-}$ in red, orange and magenta colors respectively.
		The plots correspond to the allowed parameters $M_1=1005$ GeV,
		$M_2\simeq M_1=1005.0000035$ GeV, $M_3=1385$ GeV, $m_H = 1405$
		GeV,  $Y_1= 0.4258$,  $Y_2= 0.49$ and $Y_3 = 0.66$.
	}\label{f:dental5}
\end{figure}

\begin{figure}[h!]
	\centering
	\includegraphics[height=1.55in]{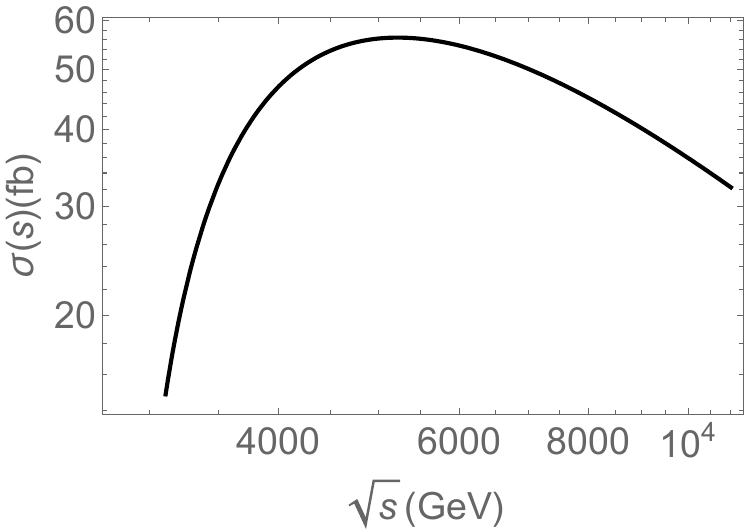}
	\caption{ Total cross section $\sigma_{e^+ e^-\to H^+ H^-}$. The
		plots correspond to the allowed parameters $M_1=1005$ GeV,
		$M_2\simeq M_1=1005.0000035$ GeV, $M_3=1385$ GeV, $m_H = 1405$
		GeV,  $Y_1= 0.4258$,  $Y_2= 0.49$ and $Y_3 = 0.66$..
	}\label{f:dental6}
\end{figure}

Subsequent sections explore the allowed parameter regions under individual and combined constraints, presented visually in figures such as Figs.~\ref{f:dental}, \ref{f:denta2}, \ref{f:dental3}, and \ref{f:dental4}. Finally, the study investigates predictions for $\sigma_{e^+ e^-\to H^+ H^-}$ cross-sections, benchmarked against constraints, with detailed contributions from $Z$, photon, and new fermions $N_{1,2,3}$ (Figs.~\ref{f:dental5} and \ref{f:dental6}).

Our analysis begins by fixing the neutrino mixing angles 
$\theta_{12}$, $\theta_{23}$, $\theta_{13}$ and the Dirac CP phase $\delta$
to their central values obtained from the latest global fit to neutrino
oscillation data~\cite{deSalas:2020pgw}. We then perform a systematic scan
over the relevant parameter space of the model, which includes the masses
of the singlet fermions $N_{1,2,3}$, the scalar masses $m_H$ and $m_0$,
and the Yukawa coupling parameters $Y_{1,2,3}$.

As discussed previously, direct collider searches impose lower bounds on
the scalar masses. In particular, for light dark matter scenarios with
$M_1 < 100~\text{GeV}$, where $N_1$ is identified as the dark matter
candidate, the parameter space is strongly constrained by lepton flavor
violation processes~\cite{Vicente:2014wga} as well as by direct searches
at the LHC~\cite{ATLAS:2019lng,CMS:2020bfa}. These constraints exclude
most of the viable region in this mass range. Consequently, we impose a
conservative lower bound of $100~\text{GeV}$ on all scalar masses.

Regarding the singlet fermions, the dark matter candidate $N_1$ remains
largely unconstrained by current collider searches. In contrast, the
masses of the heavier fermions $N_2$ and $N_3$ are required to be
sufficiently large in order to satisfy the stringent bounds from lepton
flavor violating observables, as will be demonstrated in the following.

In Fig.~\ref{f:dental}, the blue-shaded region is consistent with the
$\mu \to e\gamma$ constraint, where the remaining parameters are fixed as
$M_2 = M_1$, $M_3 = M_1 + 380$~GeV, $m_H = M_1 + 400$~GeV, $Y_2 = 0.49$
and $Y_3 = 0.66$. It is evident from the figure that satisfying the
$\mu \to e\gamma$ bound requires a heavy charged Higgs mass as well as
large masses for the fermions $N_{1,2,3}$.

In Fig.~\ref{f:denta2}, we present the allowed parameter regions after
imposing the constraint on $\frac{|\Delta m^2_{31}|}{\Delta m^2_{21}}$.
In this figure, the parameters are fixed to
$M_2 = M_1$, $M_3 = M_1 + 380$~GeV, $m_0 = M_1 + 750$~GeV, $Y_2 = 0.49$
and $Y_3 = 0.66$. The shaded regions in the $M_1$--$Y_1$ plane
are consistent with the required constraint.

We now proceed to show the allowed region in the $M_1$--$Y_1$ plane
imposed by the dark matter relic density constraint $\Omega h^2$.
The results are displayed in Fig.~\ref{f:dental3} for the parameter
choice $M_2 = M_1$, $M_3 = M_1 + 380$~GeV, $m_H = M_1 + 400$~GeV,
$m_0 = M_1 + 750$~GeV, $Y_2 = 0.49$ and $Y_3 = 0.66$. As illustrated
in the figure, larger values of the dark matter mass $M_1$ require
correspondingly larger values of the coupling $Y_1$ in order to
satisfy the relic density bound. However, this requirement is not favored by the other constraints discussed above.

It is worth emphasizing that, in the previous figures presented above,
we examined the impact of each individual constraint on the parameter
space separately. Nevertheless, the viable parameter space must satisfy
all relevant constraints simultaneously. Accordingly, in
Fig.~\ref{f:dental4}, we present the allowed region in the
$M_1$--$Y_1$ plane after imposing the dominant constraints arising from
$\mu \to e\gamma$, $\frac{|\Delta m^2_{31}|}{\Delta m^2_{21}}$, and
the dark matter relic density $\Omega h^2$. The red-colored region in
the figure simultaneously satisfies all these constraints. From this
region, we infer that viable solutions exist for which the masses of the
new particles in the model are close to or above the $1$~TeV scale.
Furthermore, this region can be used to define a set of benchmark points
that fulfill all imposed constraints on the model parameter space and
thus allow predictions for the cross section
$\sigma_{e^+ e^- \to H^+ H^-}$, which can be probed at future collider
experiments. In Fig.~\ref{f:dental5}, we show the individual contributions of the
$Z$ boson, photon, and the new singlet fermions $N_{1,2,3}$ to the cross
section $\sigma_{e^+ e^- \to H^+ H^-}$ as functions of the center-of-mass
energy $\sqrt{s}$, represented by the red, orange, and magenta curves,
respectively. The results correspond to one of the allowed benchmark
points in the parameter space, namely
$M_1 = 1005$~GeV,
$M_2 \simeq M_1 = 1005.0000035$~GeV,
$M_3 = 1385$~GeV,
$m_H = 1405$~GeV,
$Y_1 = 0.4258$,
$Y_2 = 0.49$,
and $Y_3 = 0.66$.
It is evident from the figure that the $Z$-boson contribution is the
smallest, while the contributions mediated by the singlet fermions
$N_{1,2,3}$ dominate the cross section. Finally, in Fig.~\ref{f:dental6}, we present the total cross section
$\sigma_{e^+ e^- \to H^+ H^-}$ as a function of the center-of-mass energy
$\sqrt{s}$, shown in orange, blue, and red colors, respectively. As in
the previous case, the figure corresponds to the same benchmark point in
the parameter space. As can be observed, the cross section increases
with increasing center-of-mass energy, reaches a maximum, and then
decreases as $\sqrt{s}$ continues to grow.

that satisfying $\frac{\Delta m^2_{31}}{\Delta m^2_{21}}$
constraints requires values of $Y_{1,2}$ roughly of order
${\mathcal O} (10^{-2})$ and $Y_{3}$ of order ${\mathcal O}
(10^{-1})$. On the other hand, satisfying $\mu\to e\gamma$
constraints requires values of $Y_{1,2,3}$ roughly of order
${\mathcal O} (10^{-1})$. For satisfying $\Omega \hat h^2$
constraints, values of $Y_{2,3}$ roughly of order ${\mathcal O}
(10^{-1})$ and $Y_{1}$ of order ${\mathcal O} (10^{1})$ are
required. It is known that $\mu\to e\gamma$  and $\Omega \hat h^2$
constraints can be simultaneously satisfied if the lightest
singlet fermion is almost degenerated with the next to lightest
singlet fermions. However, this possibility should be reexamined
after taking into account the strong constraints from
$\frac{\Delta m^2_{31}}{\Delta m^2_{21}}$. Finally, it is possible
to have points satisfying all constraints for masses of new
particles higher than $1$ TeV. However, their impact on other
processes will be small making them less relevant to explore new
physics scenarios.

\section{Conclusion}

In this study, we explored the process $e^+e^-\to H^+H^-$ within the framework of the scotogenic model. The contributions to the amplitude of this process arise from tree-level diagrams mediated by the photon, the Z boson, and the right-handed fermions $N_{1,2,3}$. We analyzed the processes that can impose significant constraints on the parameter space relevant to $e^+e^-\to H^+H^-$. Additionally, we evaluated the individual contributions of the photon, the Z boson, and the right-handed fermions $N_{1,2,3}$ to the cross section of $e^+e^-\to H^+H^-$, incorporating all stringent constraints on the model's parameters. Our results indicate that the dominant contribution to the cross section originates from the new singlet right-handed fermions $N_{1,2,3}$. Furthermore, we examined how the cross section varies with the center-of-mass energy for a set of benchmark points within the model's parameter space that adhere to the stringent bounds. Future $e^+e^-$ colliders will investigate the process $e^+e^-\to H^+H^-$, offering opportunities to test our predictions by either imposing stricter constraints or confirming these results.

\bibliographystyle{unsrt}

\end{document}